\documentclass[epj]{svjour}
\usepackage{graphicx}
\begin{document}
\title{Modelling Hybrid Stars in Quark-Hadron Approaches}
\subtitle{}
\author{S. Schramm\inst{1} \and V. Dexheimer\inst{2} \and R. Negreiros\inst{3}% etc
}                     % Do not remove
\offprints{}          % Insert a name or remove this line
\institute{FIAS, Ruth-Moufang-Str. 1, 60438 Frankfurt am Main, Germany \and 
Department of Physics, Kent State University, Kent, OH 44242, USA
\and
Federal Fluminense University,
Av. Gal. Milton Tavares de Souza, Gragoata, Niteroi, 24210-346, Brasil}
\date{Received: date / Revised version: date}
\abstract{
The density in the core of neutron stars can reach values of about 5 to 10 times nuclear matter
saturation density. It is, therefore, a natural assumption that  hadrons may have dissolved into quarks  under such conditions, forming a hybrid star.
This star will have an outer region of hadronic matter and a core of quark matter or even a mixed state of hadrons and quarks.
In order to investigate such phases, we discuss different model approaches that can be used in the study of compact stars as well as being applicable to a wider range of temperatures
and densities. One major model ingredient, the role of quark interactions in the stability of massive hybrid stars is discussed. In this context, possible conflicts with lattice QCD simulations are investigated.
 % end of PACS codes
} %end of abstract
\maketitle
\section{Introduction}
\label{intro}
The physics of neutron stars is one of the central areas of research in nuclear astrophysics.
Neutron stars represent a unique environment for investigating extremely dense strongly interacting matter at relatively low temperatures.
This renders neutron star physics as an ideal complement to the efforts using relativistic heavy-ion collisions in the study of matter under extreme conditions. In the latter case the fireball created in the collision zone has a very high temperature, thus sampling a different regime of QCD matter than neutron stars.  

Using equation of states of purely hadronic matter, typical maximum densities in the center of the stars  range from about 5 to 10 times nuclear matter saturation density. As these numbers correspond to close packing from a purely geometrical point of view, it is very natural to assume that at this point (and perhaps much earlier) the baryons have dissolved into their quark components. Such an object is termed hybrid star, i. e. a star with a hadronic outer core surrounding a quark, or mixed hadron-quark, inner region \cite{Glendenning1991,Glendenning1992a,Lawley2005,Pagliara2008,Dexheimer2010,COELHO2010,Lenzi2012,Orsaria2013,Endo2014,Chen2015,Li2015a}.
Depending on the equation of state used to describe the quark phase,  very different masses and radii are obtained for stars when compared to a purely baryonic star. We will discuss this issue in detail in the following sections.

\section{Modeling Compact Stars}
\label{model}
In order to investigate the properties of hybrid stars, we first introduce the formulation of the hadronic chiral mean field (CMF) model. It  is based on a SU(3) extension of a sigma-omega chiral model in a non-linear realization of chiral symmetry, including all lowest hadronic multiplets \cite{Papazoglou1998,Papazoglou1999}. The advantage of such an approach is that it also incorporates dynamic mass generation, which allows for a realistic description of chiral symmetry restoration, while being in very good agreement of nuclear and astrophysics properties.

\subsection{The Hadronic Model}
\label{hadronic}
The CMF model includes nucleons, hyperons as well as non-strange and strange mesons. As in the usual relativistic mean field approach, the baryons interact via meson fields. In addition, in the CMF approach also the baryonic masses are largely generated by the interaction with the scalar fields, except for a small explicit mass term. The specific form of the potential of the scalar fields leads to non-vanishing vacuum expectation values, thereby producing a dynamic baryon mass via the baryon-meson coupling, analogously to chiral sigma models.

The terms in the Lagrangian density relevant for the following discussions read \cite{Papazoglou1998,Papazoglou1999}
\begin{eqnarray}
\mathcal{L} = \mathcal{L}_{\rm{Kin}}+\mathcal{L}_{\rm{Int}}+\mathcal{L}_{\rm{Self}}+\mathcal{L}_{\rm{SB}}\,.
\end{eqnarray}
$\mathcal{L}_{\rm{Kin}}$ contains the kinetic energy of the fields. The interaction term $\mathcal{L}_{\rm{Int}}$ defines the linear interaction between meson fields and baryons. It reads explicitly
\begin{eqnarray}
\mathcal{L}_{\rm{Int}}&=&-\sum_i \bar{\psi_i}[\gamma_0(g_{i\omega}\omega+g_{i\phi}\phi+g_{i\rho}\tau_3\rho)+M_i^*]\psi_i~.
\label{vector}
\end{eqnarray}
The baryons $i$ interact with the vector mesons $\omega$ (isoscalar), $\rho$ (isovector), and $\phi$ (isoscalar with hidden strangeness).
The various coupling constants $g_{BM}$ are fitted as discussed below based on SU(6) symmetry \cite{Papazoglou1998}. 
The effective masses $M_i^*$ are generated by the coupling of the baryons to the scalar mesons, reading
\begin{eqnarray}
M_{i}^*&=&g_{i\sigma}\sigma+g_{i\delta}\tau_3\delta+g_{i\zeta}\zeta+M_{0_i}\,.
\end{eqnarray}
These expressions include couplings to the various scalar fields, the isosclar $\sigma$, the isovector $\delta$ and the field with hidden strangeness $\zeta$. The couplings are connected via SU(3) relations \cite{Papazoglou1998}. 

\begin{figure}
\resizebox{0.5\textwidth}{!}{
  \includegraphics[width=7cm,height=7cm]{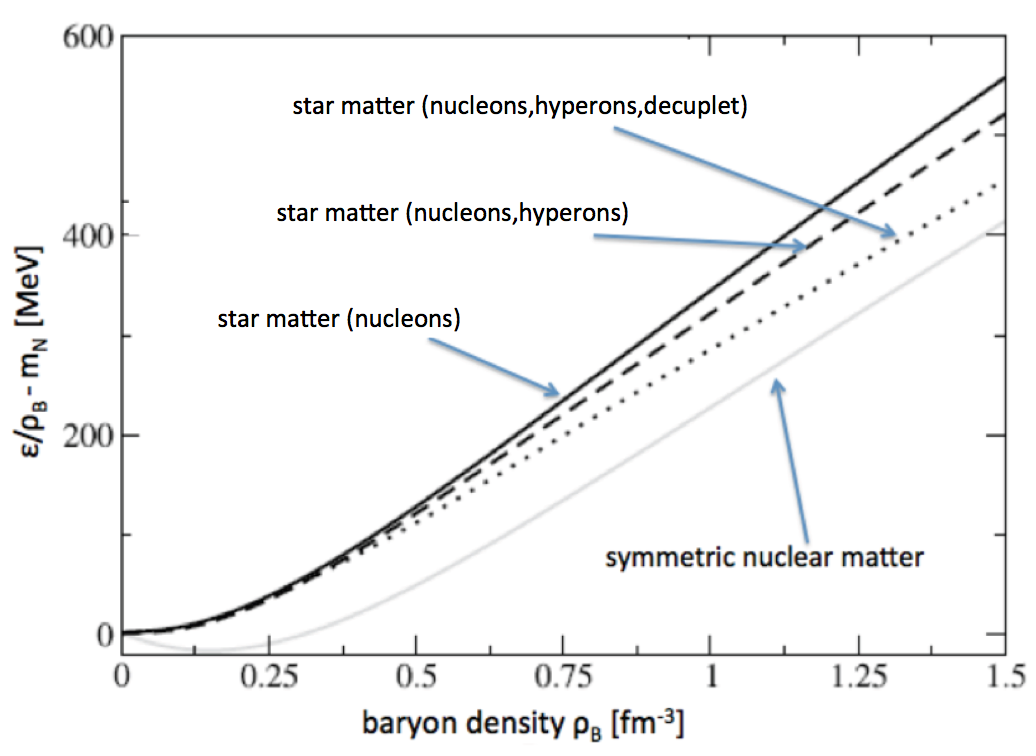}
}
%\vspace{5cm}       % Give the correct figure height in cm
\caption{Equation of state of the CMF hadronic model at T = 0. Each curve shows the equation of state for different cases: isospin symmetric nuclear matter and star matter assuming different degrees of freedom.}
\label{eosH}       % Give a unique label
\end{figure}

The scalar fields have self-interaction terms that are responsible for generating non-vanishing vacuum expectation values for
the $\sigma$ and $\zeta$ that, in quark language, correspond to the scalar vacuum condensates of non-strange and strange quarks, respectively.
Using SU(3) invariant terms for the interactions, the Lagrangian term is given by
\begin{eqnarray}
\mathcal{L}_{\rm{Self}}&=&-\frac{1}{2}(m_\omega^2\omega^2+m_\rho^2\rho^2+m_\phi^2\phi^2)\nonumber\\
&+&k_0(\sigma^2+\zeta^2+\delta^2)+k_1(\sigma^2+\zeta^2+\delta^2)^2\nonumber\\&+&k_2\left(\frac{\sigma^4}{2}+\frac{\delta^4}{2}
+3\sigma^2\delta^2+\zeta^4\right)
+k_3(\sigma^2-\delta^2)\zeta\nonumber\\
&+&k_4\ \ \ln{\frac{(\sigma^2-\delta^2)\zeta}{\sigma_0^2\zeta_0}} \nonumber \\
&+&g_4\left(\omega^4+\frac{\phi^4}{4}+3\omega^2\phi^2+\frac{4\omega^3\phi}{\sqrt{2}}+\frac{2\omega\phi^3}{\sqrt{2}}\right).
\end{eqnarray}
Finally, an explicit symmetry breaking term  that also generates masses for the pseudoscalar mesons is included
\begin{eqnarray}
\mathcal{L}_{\rm{SB}}&=&m_\pi^2 f_\pi\sigma+\left(\sqrt{2}m_k^ 2f_k-\frac{1}{\sqrt{2}}m_\pi^ 2 f_\pi\right)\zeta\,.
\end{eqnarray}
The equations of motion for hadronic matter at zero or finite temperature are determined from extremizing the grand canonical potential
\begin{eqnarray}
&\Omega / V ~=~-L_{Int}-L_{Self}-L_{SB}-L_{Vac}\nonumber &\\ &\mp T\sum_i \frac{\gamma_i}{(2 \pi)^3}
\int_{0}^{k_{F_i}} \,
d^3k \, \ln(1\pm e^{-\frac{1}{T}(E_i^*(k)-\mu_i^*)}),&
\end{eqnarray}
including a heat bath of quasiparticles.
This model approach has been used in many applications in relativistic heavy-ion physics \cite{Steinheimer2010,Steinheimer2008}, nuclear structure calculations \cite{Schramm2002a,Schramm2015}, as well as astrophysical studies \cite{Dexheimer2008b,Schurhoff2010a}.
Fig. \ref{eosH} shows a sample of these equations of state at vanishing temperature \cite{Dexheimer2008b}. In-between the equation of state of nucleonic matter and the relatively soft equation of state for isospin symmetric matter, the figure also incorporates results taking into account the hyperon states  as well as an extension of the calculation that also includes the spin-3/2 decuplet of baryon resonances. As can be observed, adding further degrees of freedom leads to a softening of the equation of state \cite{Dexheimer2008b,Schurhoff2010a}. Note that for the isospin symmetric nuclear matter calculation net strangeness is set to zero, whereas in general stellar matter is calculated assuming beta equilibrium and charge neutrality.

\subsection{Hybrid stars}
\label{quarks1}
As has been argued in the introduction, it is very natural to assume that, at some density, quark degrees of freedom appear inside compact stars. Before discussing the extension of the hadronic model to include quarks, let us look at some general aspects of hybrid star solutions.
For this purpose, we combine an often-used hadronic model (G300 \cite{Glendenning1989}) with a simple MIT bag quark equation of state  \cite{Negreiros2012,Negreiros2010}. 
In this approach we can adjust the  quark phase bag pressure $B$. The choice of $B$ essentially determines at what chemical potential the phase transition to quark matter occurs and, connected to this, it determines the strength of the first-order phase transition.

\begin{figure}
\resizebox{0.5\textwidth}{!}{
  \includegraphics{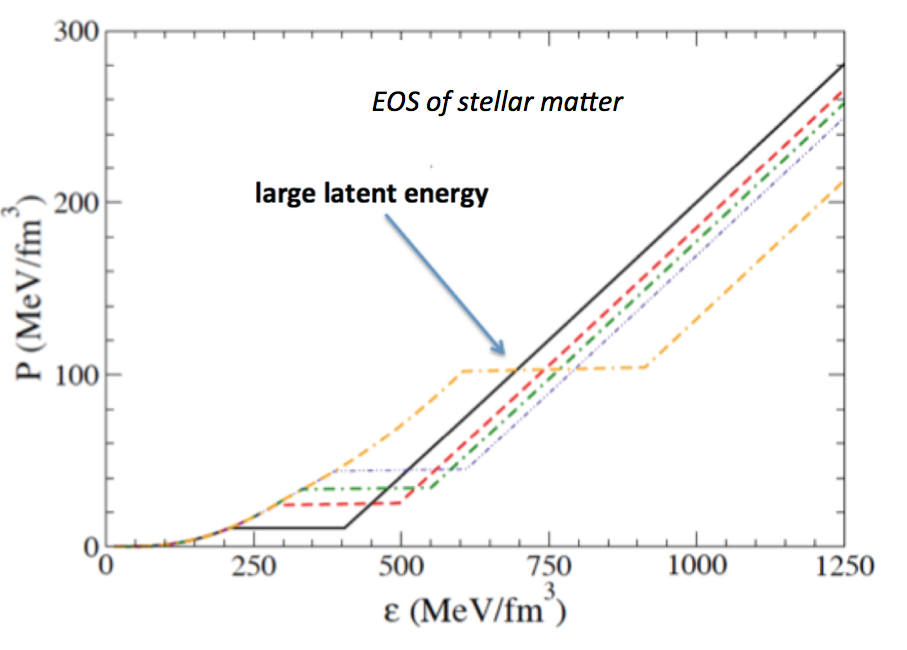}
}
%\vspace{5cm}       % Give the correct figure height in cm
\caption{Equation of state of the G300/MIT hybrid model. The various curves correspond to different choices of the vacuum pressure with values $B^{1/4} =  160, 165, 167.5, 170, 180\,$MeV (larger values generating later transitions).
}
\label{MITeos}       % Give a unique label
\end{figure}
\begin{figure}
\resizebox{0.5\textwidth}{!}{
  \includegraphics{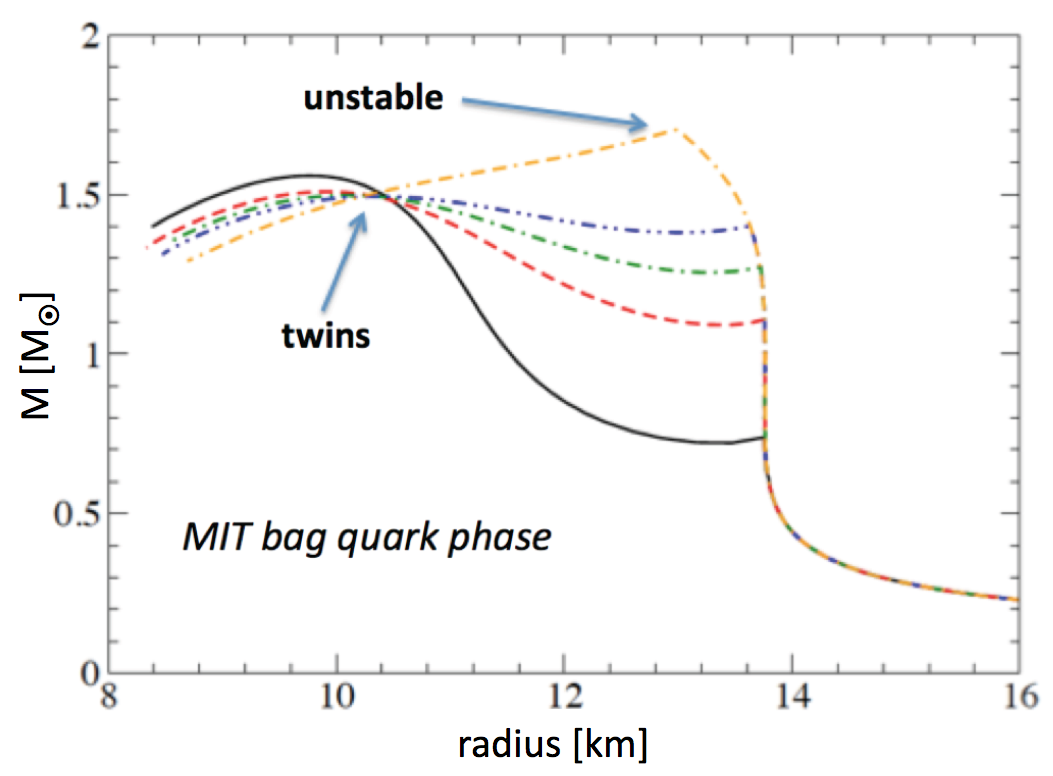}
}
%\vspace{5cm}       % Give the correct figure height in cm
\caption{TOV solutions for the equations of state shown in Fig. \ref{MITeos}. Strong first-order phase transitions with large latent energies lead to stellar instability  when quarks appear.}
\label{MIT}       % Give a unique label
\end{figure}

Fig. \ref{MITeos} shows that for larger values of the bag pressure the transition point is shifted to higher densities and the jump in the energy density also significantly increases. Both of these have direct consequences for hybrid star solutions. The curves in Fig. \ref{MIT} are the solutions of the 
Tolman-Oppenheimer-Volkov (TOV) equations for static spherically symmetric stars \cite{Tolman1939,Oppenheimer1939} for this set of equations of state. The kinks in the curves in Fig. \ref{MITeos} mark the onsets of the quark phase. The general effect of this instability is a reduction of the maximum star masses as well as a shift of the radius of these stars to smaller values.  
In the case of very strong transitions, the onset can even lead to a removal of all stars with a quark core from the range of stable solutions. This is the case for the equation of state generated with a bag pressure $B = (180 $MeV$)^4$. 
For some cases, also two maxima develop, which would correspond to twin star solutions, meaning two branches of stable hadronic and hybrid stars, respectively \cite{Gerlach1968,Glendenning1998,Schaffner-Bielich2002a,Dexheimer2015,Benic2014}. 

\begin{figure}
\resizebox{0.5\textwidth}{!}{
  \includegraphics[width=7cm,height=6cm]{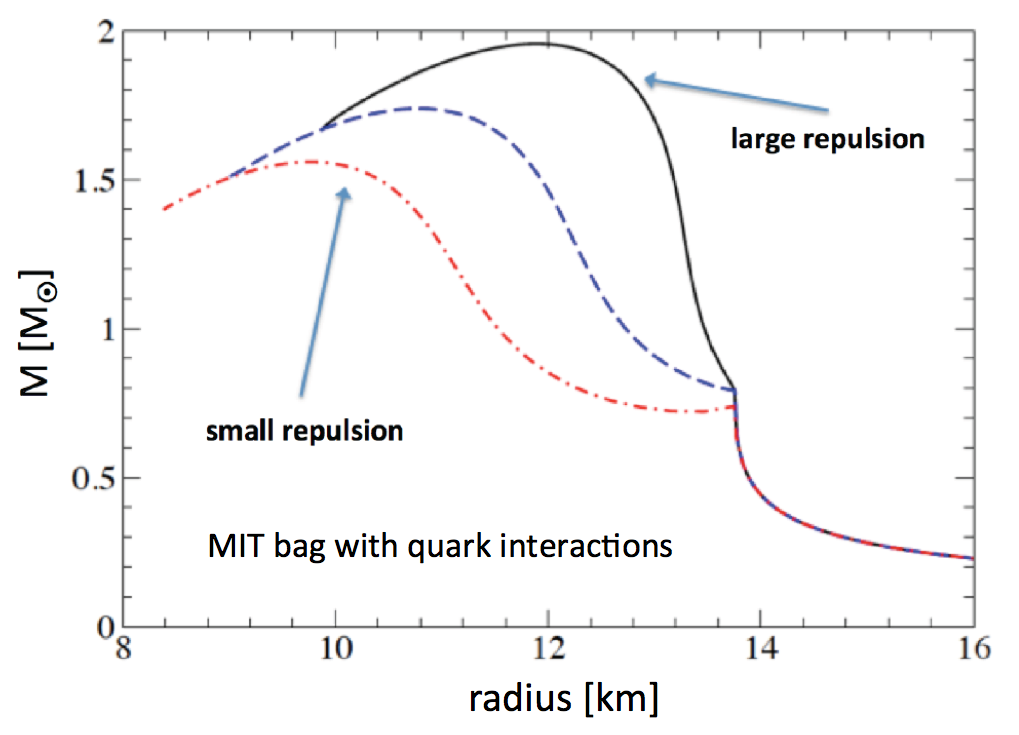}
}
%\vspace{5cm}       % Give the correct figure height in cm
\caption{As Fig. \ref{MIT}, but effectively including a repulsive interaction to the quark equation of state. Results for coupling strengths
$\alpha = 0.1, 0.3, 0.7$ are shown. Large repulsion
leads to larger star masses and weaker phase transitions.}
\label{MITrep}       % Give a unique label
\end{figure}

This general behaviour can be modified by including interactions between the quarks as shown in Fig. \ref{MITrep}.
Following the simplified procedure outlined in Ref. \cite{Alford2005}, the modified quark phase is described by the grand canonical potential
\begin{equation}
\Omega_q = - \frac{3}{4 \pi^2} \mu^2 [ (1- 2 \alpha_s / \pi) \mu^2 - m_s^2 ]+ B ~~.
\end{equation} 
Here, the additional term proportional to the strong interaction coupling strengh $\alpha_s$ contains the repulsive interaction between the quarks.
As was pointed out in Ref. \cite{Alford2005}, adjusting the interaction strength accordingly, the transition between quarks and hadrons becomes weak, effectively leading to similar masses and radii for the hadronic and hybrid stars.
In this way, a strong repulsive quark interaction helps to stabilize hybrid stars.
A general discussion of the stability of hybrid stars, using model-independent parameterized equations of state, can be found in Ref. \cite{Alford2013a}.
 However, this procedure of stabilizing hybrid stars creates different problems, as it will be briefly discussed in Section \ref{lattice}.

\subsection{Unifying hadron and quark models}
\label{model1} 

\begin{figure}
\resizebox{0.5\textwidth}{!}{
  \includegraphics[width=7cm,height=5cm]{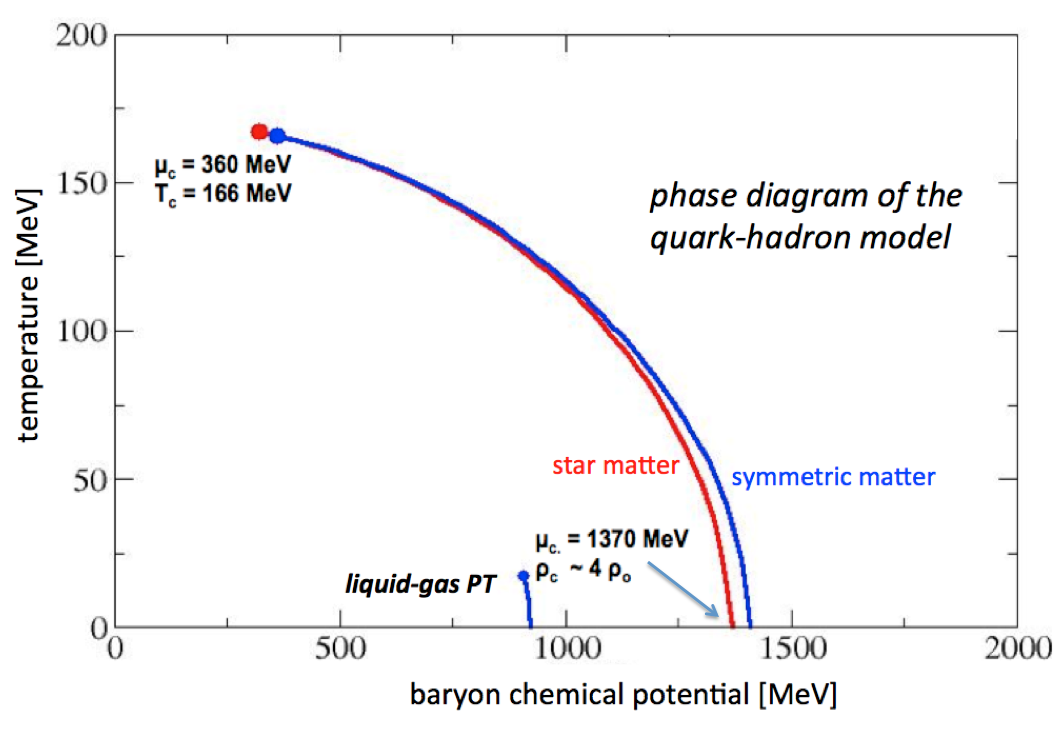}
}
%\vspace{5cm}       % Give the correct figure height in cm
\caption{Phase diagram of the quark-extended CMF model. Apart from the nuclear liquid-gas first-order phase transition, the chiral/deconfinement transition exhibits a critical end-point at small chemical potentials. Results for isospin symmetric and star matter are shown.}
\label{qhpd}       % Give a unique label
\end{figure}

As the model approach under discussion is supposed to be (and has been) applied to a large range of temperatures and densities, including conditions that exist in the fireball of heavy-ion collisions, its general phase structure should be in accordance with the known features of the QCD phase diagram. The well-established properties are (i) bound nuclear matter, i.e. a first-order liquid-gas phase transition at the nuclear matter saturation density of about 0.15 fm$^{-3}$ and (ii) a smooth crossover for vanishing net baryon density at a temperature of about 160 MeV as it has been unambiguously established in lattice QCD simulations \cite{Borsanyi2014,Bazavov2014}. Thus, a simple connection of two separate models of the hadronic and the quark phases, as it was used in the previous section, is not sufficient to fulfil these requirements. Connecting both equations of state will necessarily lead to a first-order transition for all conditions (except perhaps for a single point, i.e. chemical potential, by fine-tuning parameters). Therefore, a unified model including hadronic and quark degrees of freedom is required. In \cite{Dexheimer2010} such an approach has been developed for the first time (see also Refs. \cite{Turko2014,Benic2015}. In this formulation, quarks and baryons interact via the scalar and vector meson condensates. In addition, in analogy to the quark Polyakov loop - NJL (PNJL) model, an effective field $\Phi$ is introduced that effectively describes the order parameter for the deconfinement transition \cite{Ratti2006a,Fukushima2004}.
This field has a potential energy term, which is fitted in order to reproduce the thermodynamic and Polyakov loop measurements in lattice QCD simulations.
A simplified potential of this type was proposed in Ref. \cite{Dexheimer2010}
\begin{eqnarray}
&U=(a_0T^4+a_1\mu^4+a_2T^2\mu^2)\Phi^2&\nonumber\\&+a_3T_0^4\log{(1-6\Phi^2+8\Phi^3-3\Phi^4)}.&
\label{potpol}
\end{eqnarray}
The constants $a_0$ and $a_3$ were fitted to obtain a reasonable agreement with lattice results at zero chemical potential. 
The potential term $U$ also contains terms that depend on the chemical potential. They can be adjusted to generate a first-order phase transition at zero temperature to a deconfined state at a chosen baryon chemical potential. Further, they can be adjusted to lead to a phase diagram with a critical end-point of the first-order phase transition line that is in agreement with a particular lattice QCD calculation \cite{Fodor2004}.
The resulting phase diagram of such an approach is shown in Fig. \ref{qhpd}. The figure exhibits a transition line with the end-point at small chemical potential. It also shows the nuclear liquid-gas phase transition. The two slightly shifted lines mark the difference of the transition
for isospin-symmetric and  star matter. 

Another ingredient of this model approach is the coupling of the $\Phi$ field to the masses of the baryons and quarks. This is a simple phenomenological way of ensuring that, for small value of the field, quarks are not populated and vice versa in the case of baryons. The 
effective masses read explicitly (for details, see Ref. \cite{Dexheimer2010}):
\begin{eqnarray}
&M_{B}^*=g_{B\sigma}\sigma+g_{B\delta}\tau_3\delta+g_{B\zeta}\zeta+M_{0_B}+g_{B\Phi} \Phi^2,&
\label{6}
\end{eqnarray}
\begin{eqnarray}
&M_{q}^*=g_{q\sigma}\sigma+g_{q\delta}\tau_3\delta+g_{q\zeta}\zeta+M_{0_q}+g_{q\Phi}(1-\Phi).\nonumber&\\&
\label{7}
\end{eqnarray}

\begin{figure}
\resizebox{0.5\textwidth}{!}{
  \includegraphics[width=7cm,height=5cm]{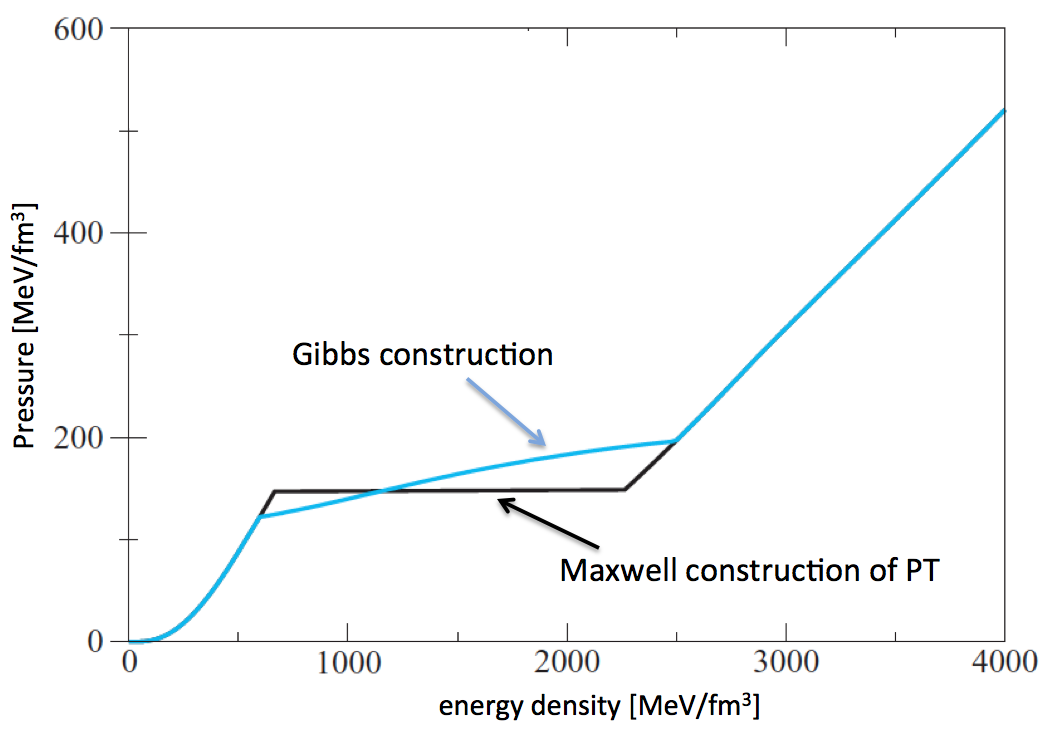}
}
%\vspace{5cm}       % Give the correct figure height in cm
\caption{Equation of state for star matter in the quark-extended CMF model at zero temperature. Two assumptions, local (Maxwell) and global (Gibbs) charge neutrality, are considered.}
\label{qheos}       % Give a unique label
\end{figure}

With this model in hand, it is now possible to calculate hybrid stars as well as quark-hadron matter at high temperatures and/or densities in general.
Using the model parameters as given in Ref. \cite{Dexheimer2010,Negreiros2010a}, the equation of state of star matter can be determined. The result for vanishing temperature is shown in Fig. \ref{qheos}. As can be seen in this figure, the equation of state has a strong first-order phase transition.
The figure includes two different treatments of the transition: first, local charge neutrality in both phases is assumed, which corresponds to a Maxwell construction, and a transition that happens at a fixed pressure. If one relaxes this condition to a global charge neutrality, with opposite  charges in  quark and hadronic phases, one obtains a range with a mixed phase and varying pressure. Which of these scenarios is correct, depends on the QCD surface tension which has to be taken into account in a mixed phase with bubbles of different phases. Unfortunately the value of the surface tension, not to mention its variation with density, is largely uncertain \cite{Pinto2012,Mintz2013}.

\begin{figure}
\resizebox{0.5\textwidth}{!}{
  \includegraphics[width=7cm,height=7cm]{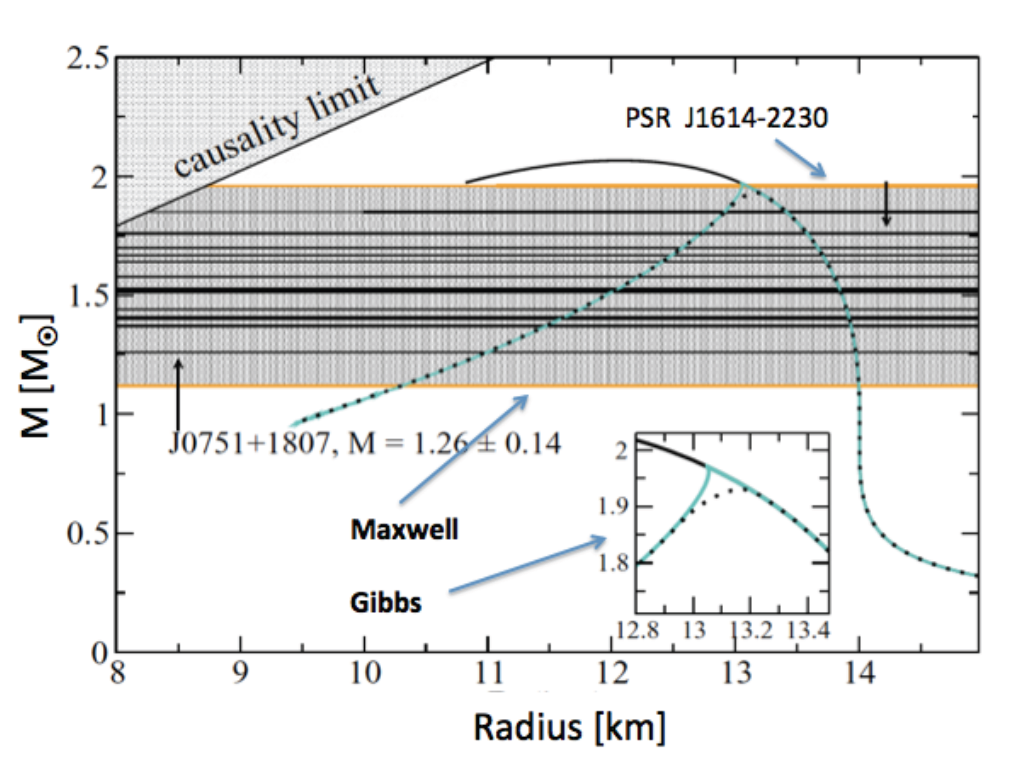}
}
%\vspace{5cm}       % Give the correct figure height in cm
\caption{Mass-radius diagram of the quark-hadron model from Section \ref{model1}. The strong first-order phase transition leads to unstable hybrid stars using Maxwell construction. With the Gibbs procedure, hybrid stars with a core of  a mixed state of quarks and baryons become stable.}
\label{MRhybrid}       % Give a unique label
\end{figure}

\begin{figure}
\resizebox{0.5\textwidth}{!}{
  \includegraphics[width=7cm,height=7cm]{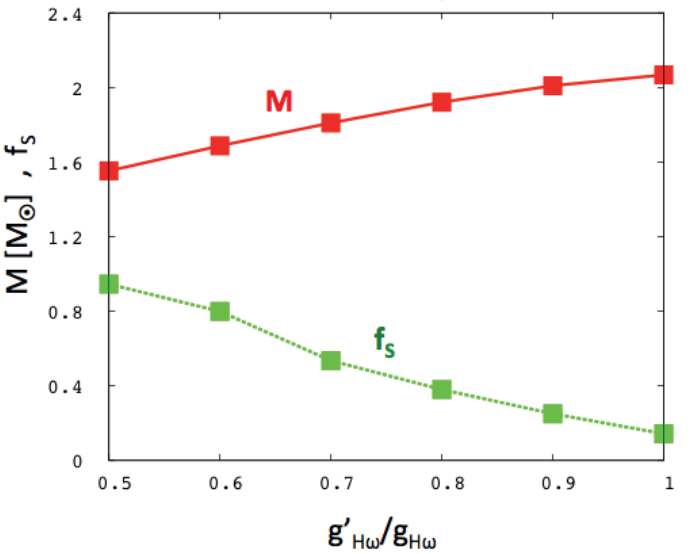}
}
%\vspace{5cm}       % Give the correct figure height in cm
\caption{Stellar mass and central strangeness fraction $f_s$ as function of the hyperon-vector meson interaction strength.
Reducing the coupling substantially increases the strangeness fraction and reduces star masses.}
\label{fs}       % Give a unique label
\end{figure}

The equations of state from Fig. \ref{qheos} can then be used to solve the TOV equations. The obtained results are presented in Fig. \ref{MRhybrid}.
The full line in the large figure is the result for stars without taking into account quarks but including hyperons.  The main reason that the hadronic star including hyperons (hyper star) still has a large maximum mass, which is only reduced by about $0.06 M_\odot$  \cite{Dexheimer2008b,Dexheimer2015a}, arises from the fact that in this full flavor-SU(3) approach the hyperons experience a strong repulsive force from $\phi$ and $\omega$ meson exchange.
As an illustration of this feature, Fig. \ref{fs} shows the resulting maximum mass and the strangeness content $f_s$  (defined as the average strangeness per baryon) in the center of the star as function of the scaled hyperon-omega meson interaction (note that one could alternatively also scale the interaction with the $\phi$ meson). Decreasing artificially the coupling shows a substantial increase in strange particles and the accompanying drop in maximum mass.

Going back to the discussion of figure \ref{MRhybrid},  including the quarks in the calculation, the resulting hybrid star solution features a kink at around 2 solar masses. As we discussed in more general terms in Section \ref{quarks1}, the latent energy in the phase transition adds a large amount of energy density to the star matter when quark states begin to be  populated, but not much more pressure. Therefore, using the Maxwell-constructed equation of state, the hybrid star becomes unstable with respect to collapse. On the other hand, using the Gibbs construction, a mixed phase of quarks and baryons survives in the core of the star as is indicated in the inset of the figure. This region can extend up to 2 km in the center of the star.

\begin{figure}
\resizebox{0.5\textwidth}{!}{
  \includegraphics[width=7cm,height=7cm]{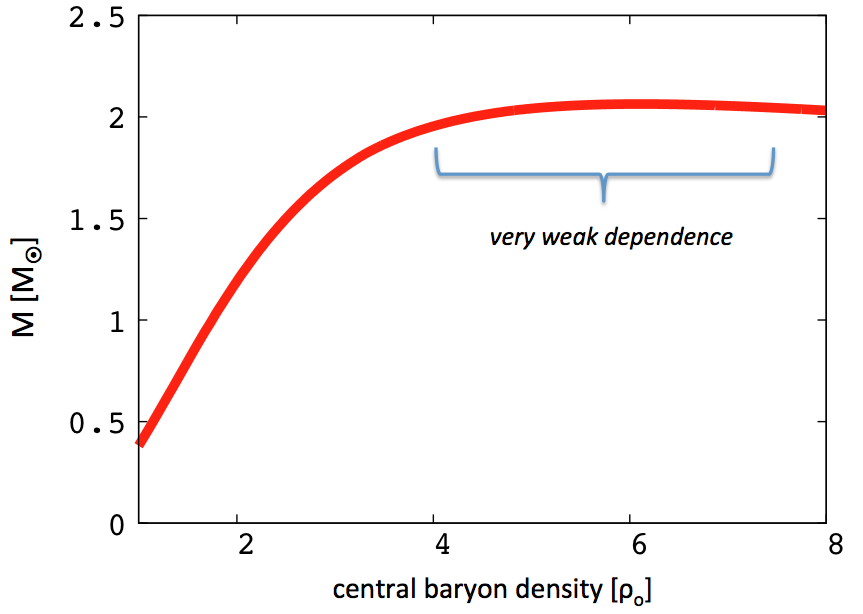}
}
%\vspace{5cm}       % Give the correct figure height in cm
\caption{Star masses as function of their central density $\rho_c$ for the hadronic CMF model. As can be seen for masses around 2 solar masses,
the actual star mass is relatively insensitive to $\rho_c$ in this region.}
\label{Medens}       % Give a unique label
\end{figure}

It is interesting to note that, although the maximum central density shifts from about $6 \rho_0$ to $4 \rho_0$ (in the Maxwell case) due to the instability introduced by the quark phase there is only a moderate reduction in maximum mass of about 0.1 solar masses. The reason for this behaviour is illustrated in Fig. \ref{Medens}. Here, the star masses are plotted as function of central density for the purely hadronic stars. The figure shows that the mass is quite insensitive to the central density for a wide range of densities below the maximum mass value. Therefore, in this case even a reduction of the value of the central density by about 30 percent does not appreciably change the maximum star mass.

\subsection{Excluded volume corrections}
\label{exvol}

\begin{figure*}
\resizebox{\textwidth}{!}{
\centerline{
  \includegraphics[width=13cm,height=7cm]{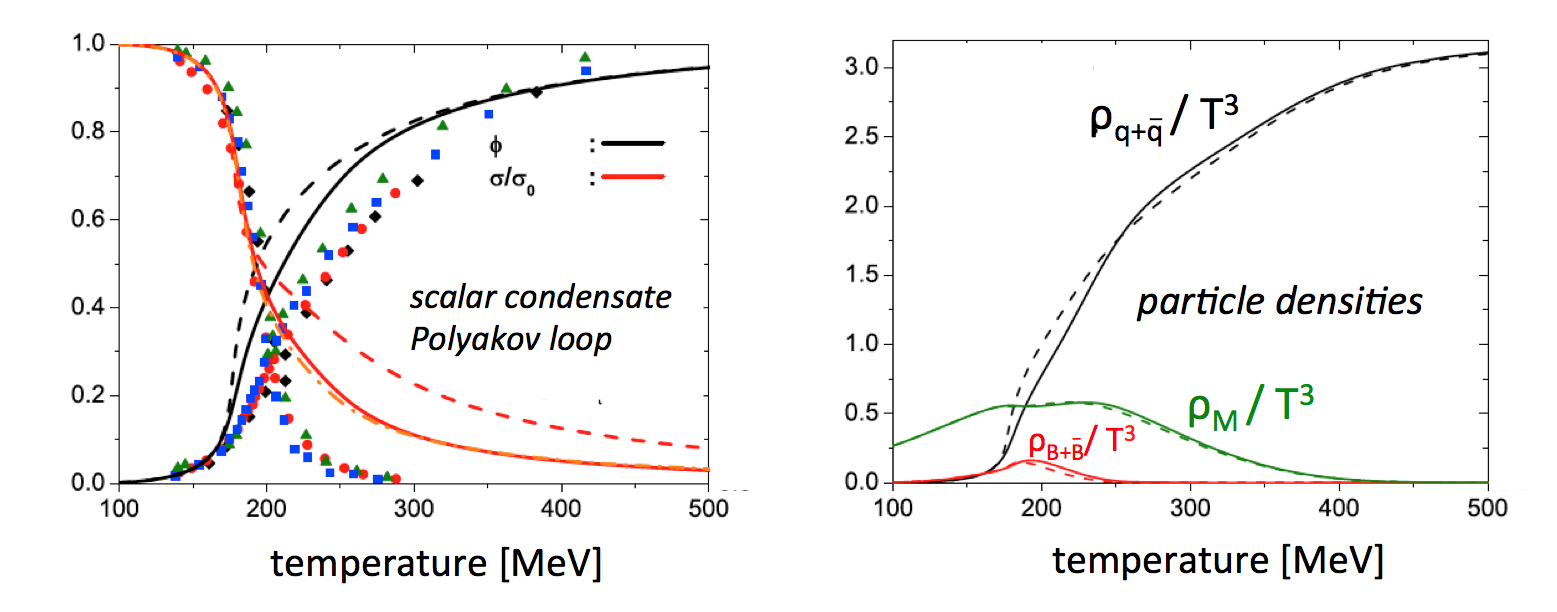}
}
}
\caption{Model results including excluded volume corrections  at vanishing chemical potential. The figure contains plots of the temperature dependence of the effective Polyakov loop field $\Phi$, the scalar field
$\sigma$, and the densities of quarks and hadrons with data points from lattice calculations \cite{Bazavov2009}.}
\label{sigmaphi}       % Give a unique label
\end{figure*}

In the development of the quark-hadron approaches as discussed in this article, a second ansatz 
that incorporates a different mechanism in transiting from baryons to quarks at high densities and temperatures was developed.
The hadronic model used is the same as the one discussed in Section \ref{hadronic}. 
In the  calculation presented here, the expression for the Polyakov loop potential is the one used 
 in a number of PNJL calculations
\cite{Ratti2006a,Fukushima2004}:
\begin{eqnarray}
U&=&-\frac12
a(T)\Phi\Phi^*\nonumber\\&+&b(T)ln[1-6\Phi\Phi^*+4(\Phi^3\Phi^{*3}
)-3(\Phi\Phi^*)^2]~.
\end{eqnarray}
The particular logarithmic term, which is also present in Eq. (\ref{potpol}), corresponds to the Haar measure of integrating over the SU(3) Polyakov loop matrix. The temperature-dependent couplings are defined as $a(T)=a_0 T^4+a_1 T_0 T^3+a_2 T_0^2 T^2$ and $b(T)=b_3 T_0^3 T$. 
The numerical values of the various parameters are adopted from Ref. \cite{Ratti2006a}. This choice ensures a correct first-order
phase transition in the pure gauge sector, i.e. for the theory including only the $\Phi$ field. 
Again, following the spirit of the PNJL model, the distribution functions of the quarks and antiquarks contain the Polyakov loop field
\begin{equation}
	\Omega_{q}=-T \sum_{j\in Q}{\frac{\gamma_i}{(2 \pi)^3}\int{d^3k \ln\left(1+\Phi \exp{\frac{E_i^*-\mu_i}{T}}\right)}}~~,
\end{equation}
and
\begin{equation}
	\Omega_{\overline{q}}=-T \sum_{j\in Q}{\frac{\gamma_i}{(2 \pi)^3}\int{d^3k \ln\left(1+\Phi^* \exp{\frac{E_i^*+\mu_i}{T}}\right)}}~~,
\end{equation}
where $\gamma_i$ stands for the spin and color degrees of freedom.
This ensures suppression of quarks in the confined state ($\Phi = 0$).
In the case of hadrons, an excluded volume term is introduced that effectively mimics short-range repulsion of overlapping hadrons.
Here, we follow a very simple approach, assuming different excluded volumes for quarks, baryons, and mesons of the form
\begin{eqnarray}
 v_{Quark}&=&0 \, , \nonumber \\
 v_{Baryon}&=&v \, , \nonumber \\
 v_{Meson}&=&v/a \, , \nonumber \\
\end{eqnarray}
with a value $v = 0.64\,$fm$^3$.
The parameter $a$ takes into account that the mesonic volume is smaller than the one for baryons.
We assume it to be $a=8$, which implies a meson radius of half the size of the baryonic one.
In this basic implementation, we neglect possible Lorentz contraction effects for the excluded volumes, which were discussed in Ref. \cite{Bugaev2000,Bugaev2008}, or density or temperature-dependent  hadron sizes \cite{Kapusta1983,Satarov2015}.
For thermodynamic consistency,
the modified chemical potential $\widetilde{\mu}_i$ of particle species $i$ is given by
\begin{equation}
	\widetilde{\mu}_i=\mu_i-v_{i} \ P~~,
\end{equation}
where $P$ is the sum over all partial pressures. In this formalism the thermodynamic quantities are calculated with respect to the temperature $T$ and the modified chemical potentials $\widetilde{\mu}_i$. Thermodynamic consistency is fulfilled by multiplying the densities  (energy density $\widetilde{\epsilon_i}$, number density  $\widetilde{\rho_i}$, entropy density $\widetilde{s_i}$), calculated with the values $\widetilde{\mu}_i$  by a factor $f$, which is the ratio of the total volume $V$ to the unoccupied reduced volume $V'$
\begin{equation}
	f=\frac{V'}{V}=(1+\sum_{i}v_{i}\rho_i)^{-1}~~.
\end{equation}
Thus, for instance, the energy density then reads
\begin{eqnarray}
\epsilon&=&\sum_i f \ \widetilde{\epsilon_i} ~~.
\end{eqnarray}
 Some results of this model for vanishing baryon chemical potential are collected in Fig. \ref{sigmaphi}.
 The plots compare model results to lattice QCD calculations  \cite{Bazavov2009}. The left panel shows the temperature dependence of the Polyakov loop and the scalar condensate, whereas the right panel depicts the densities of quarks and hadrons (summing particle and anti-particle densities). As the transition in this regime is a smooth crossover, it is rather natural to assume that there is a mixed state, and not a mixture of two separate phases, of hadrons and quarks as it is observed in this model.

\begin{figure}
\resizebox{0.5\textwidth}{!}{
  \includegraphics[width=7cm,height=7cm]{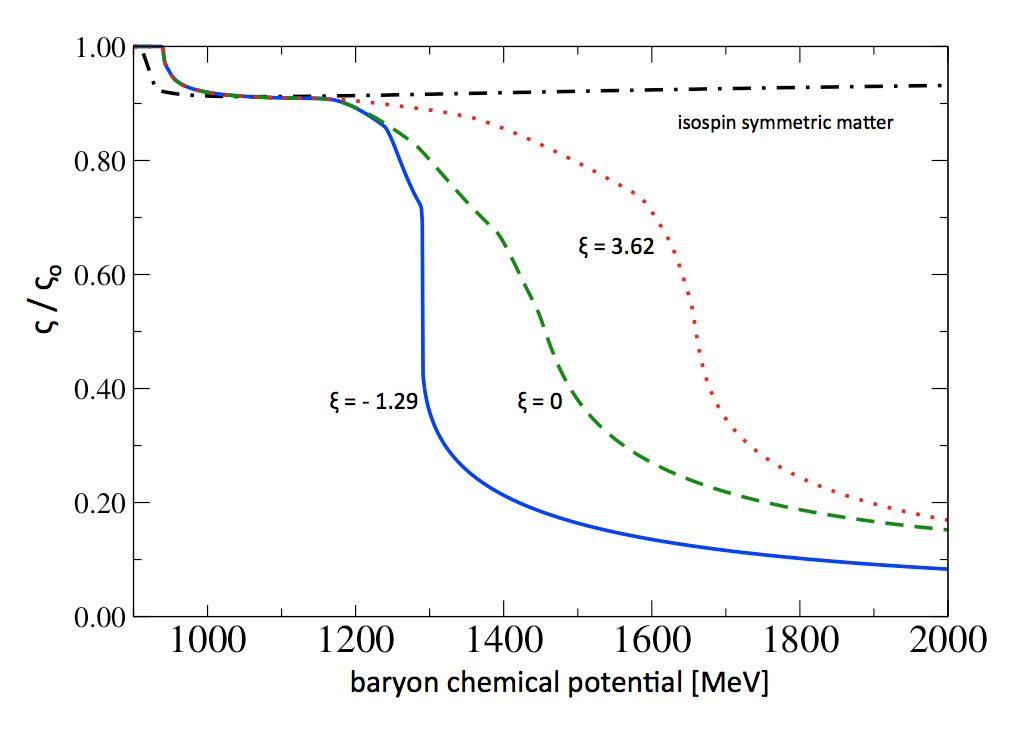}
}
%\vspace{5cm}       % Give the correct figure height in cm
\caption{Scalar field $\zeta$ with hidden strangeness normalized by its vacuum value. Depending on the quark interaction strength $\xi$ a first-order transition might occur for star matter.}
\label{zeta}       % Give a unique label
\end{figure}

\begin{figure}
\resizebox{0.5\textwidth}{!}{
  \includegraphics[width=7cm,height=7cm]{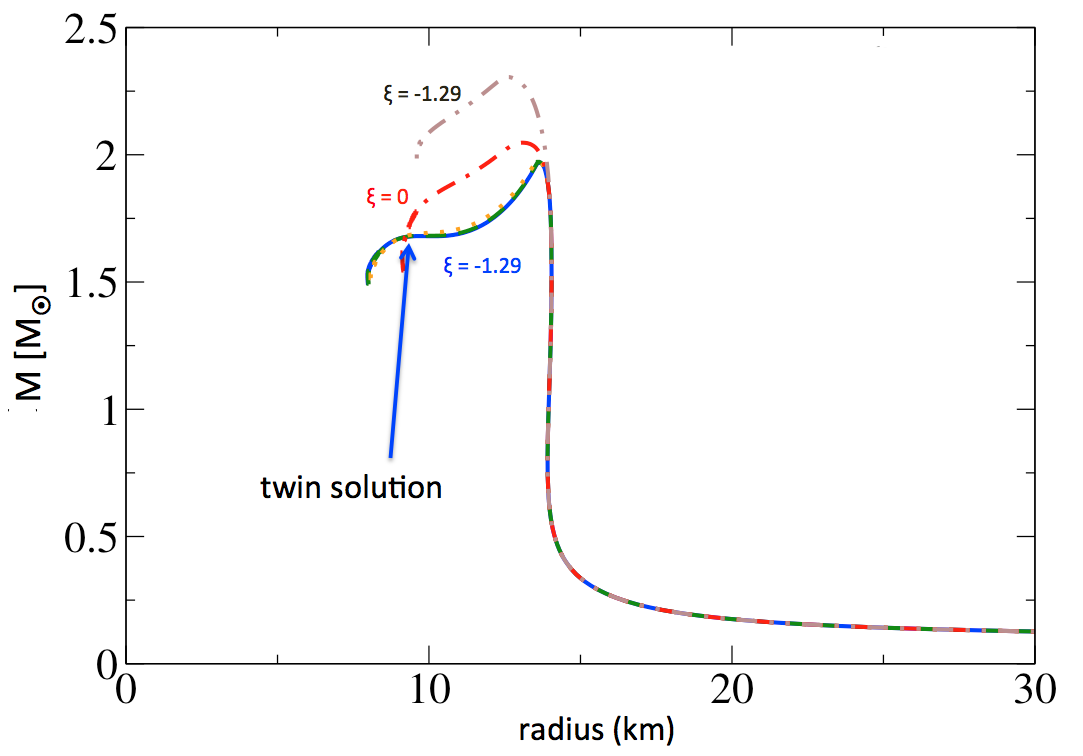}
}
%\vspace{5cm}       % Give the correct figure height in cm
\caption{TOV solutions for  hybrid stars using different strengths of the quark-$\omega$ and quark-$\phi$ couplings.
For specific choices, a twin-star solution might occur.}
\label{twin}       % Give a unique label
\end{figure}

\begin{figure}
\resizebox{0.5\textwidth}{!}{
  \includegraphics[width=7cm,height=7cm]{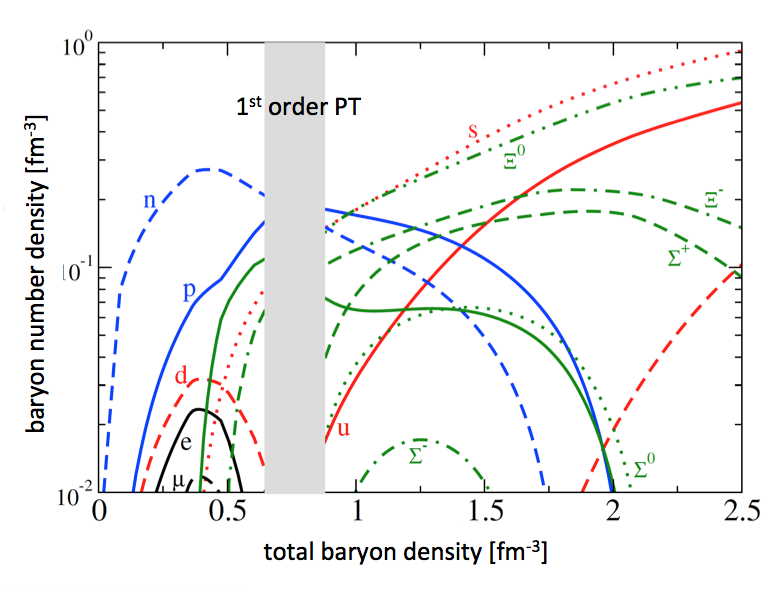}
}
%\vspace{5cm}       % Give the correct figure height in cm
\caption{Baryon number densities for hybrid star matter including a first-order phase transition ($\xi = -1.29$). One can observe the increased strangeness in the high-density phase.}
\label{twinpop}       % Give a unique label
\end{figure}

When calculating the phase structure of isospin symmetric matter for a wide range of temperatures and densities, this model does not lead to a first-order phase transition (in addition to the liquid-gas phase transition). This feature does not violate any established constraints, as the type of phase transition of matter at high density is unknown. However, in the case of charge-neutral matter in beta equilibrium, the situation is different. More precisely, some solutions for different interaction strengths of the non-strange quarks to the $\omega$ field, $g_{q\omega}$, and of the strange quark to the corresponding $\phi$ field, $g_{s\phi}$,
(with the relative strength $\xi = g_{s\phi} / g_{q\omega}$) produce a first-order transition to a strangeness-enriched phase for star matter. This is shown in Fig. \ref{zeta} for the strange scalar field $\zeta$. The corresponding solutions for the star families shown in Fig. \ref {twin} demonstrate that such a transition might lead to a small range of second family ``twin'' stars (see also Ref. \cite{Benic2014} for the case of very heavy twin solutions). Looking at the different particle densities as is shown in Fig. \ref{twinpop}, one can see a steep rise in the s quark population beyond the phase transition. Note that even stars that do not reach the critical density of the phase transition already contain quarks, being  also hybrid stars.

Twin star solutions have recently been revisited in the literature as they pose a particular solution for the so-called ``hyperon puzzle''. As the appearance of quark matter suppresses the amount of hyperons at high density, this scenario might allow for small, but massive stars.  


One very interesting and promising venue in the study of compact stars is the investigation of their thermal evolution.
This evolution is strongly dependent on the particle composition of the star, which makes it an ideal tool to gather information
on the actual particle content of the star. 
The equations that describe the stellar thermal evolution in a general relativistic setting read
\cite{Weber1999}
\begin{eqnarray}
  \frac{ \partial (l e^{2\phi})}{\partial m}& = 
  &-\frac{1}{\rho \sqrt{1 - 2m/r}} \left( \epsilon_\nu 
    e^{2\phi} + c_v \frac{\partial (T e^\phi) }{\partial t} \right) \, , 
  \label{coeq1}  \\
  \frac{\partial (T e^\phi)}{\partial m} &=& - 
  \frac{(l e^{\phi})}{16 \pi^2 r^4 \kappa \rho \sqrt{1 - 2m/r}} 
  \label{coeq2} 
  \, .
\end{eqnarray}
Here, the variables $r$, $\rho(r)$ and $m(r)$ denote the radial distance from the center of the star, the
energy density, and the stellar mass, respectively. The various thermal properties that enter the evolution are  the temperature distribution $T(r,t)$, luminosity $l(r,t)$, the neutrino emissivity $\epsilon_\nu(r,T)$, thermal conductivity
$\kappa(r,T)$ as well as the specific heat $c_v(r,T)$. The equations have to be supplemented by the appropriate boundary conditions. These are given by demanding a vanishing heat flux at the center of the star and fixing the luminosity at the surface. Here, one has to take care of the relation between the temperature of the mantle and the actual photosphere as described in \cite{Gudmundsson1982,Gudmundsson1983}. 
The calculation takes into account all standard cooling channels via neutrino emission, including direct and modified Urca processes as well as neutrino bremsstrahlung and the pair-breaking-formation process that occurs during the onset of pairing. The pairing channels include neutron singlet ($^1 S_0$) in the crust, neutron triplet ($^3 P_2$) in the core as well as proton singlet ($^1 S_0$) in the core. In addition to that one may also have to consider the different quark pairing that may occur in the deconfined quark phase. Among the most promimnnt quark pairing patterns are the Color-Flavor-Locked (CFL) and the 2-color (2SC) superconductivity.

\begin{figure}
\resizebox{0.5\textwidth}{!}{
  \includegraphics[width=7cm,height=7cm]{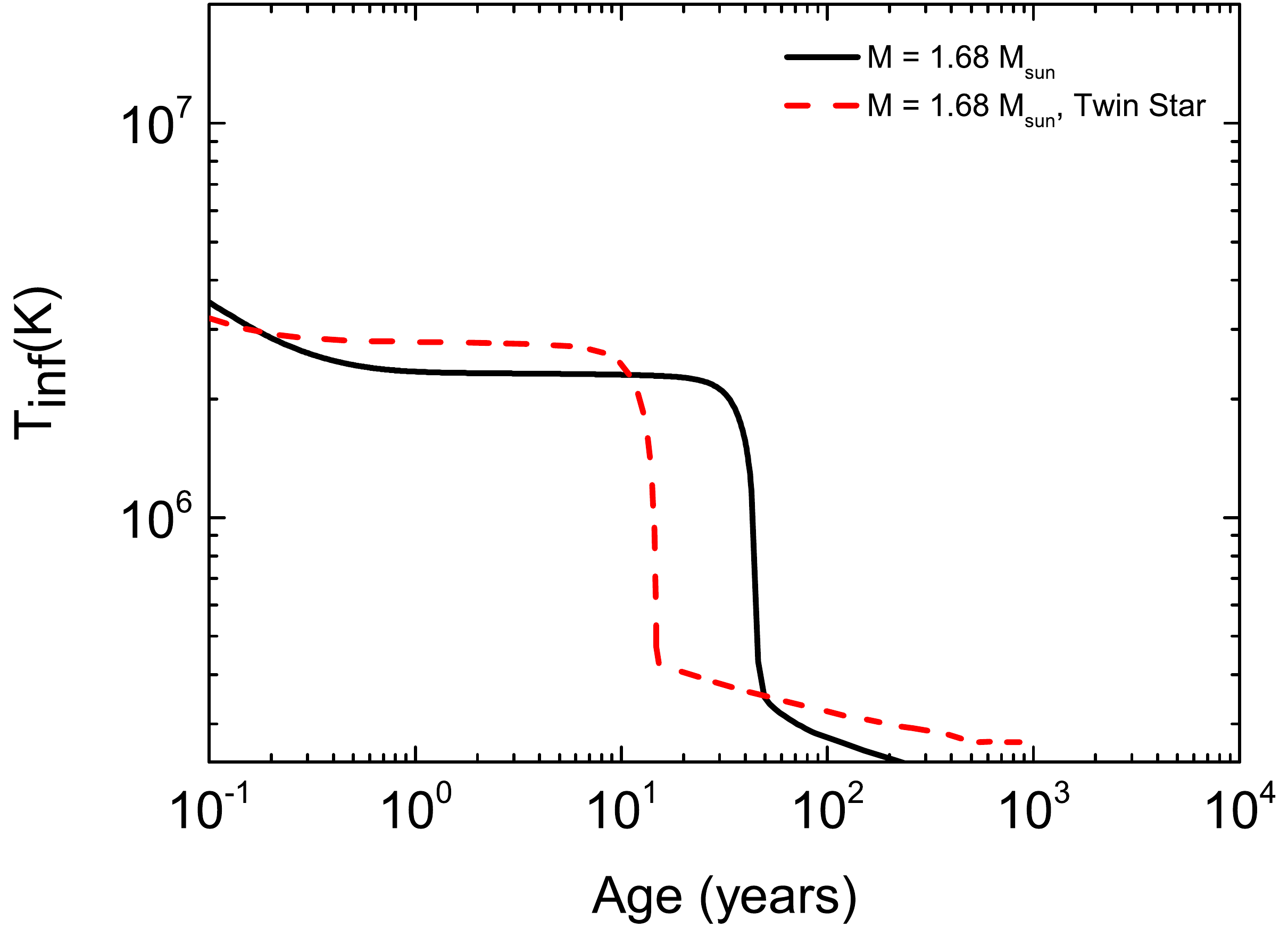}
}
\caption{Cooling curve of the hybrid star and its high-strangeness twin solution with the same mass of 1.68 M$_\odot$.
As can be seen in the figure, the twin star cools significantly faster.}
\label{twin_cool}       % Give a unique label
\end{figure}

As an example of such a study, Fig. \ref{twin_cool} shows the cooling curve of 
the two equal mass stars, one from the normal branch and one from the twin branch, which are shown in  Fig. \ref{twin}.
Although both stars have equal mass of $M = 1.68\, M_\odot$, their thermal evolution is distinctly different from each other. As can be seen, the strangeness-rich twin star cools significantly faster than the star from the ordinary branch.
This might help to distinguish stars that might have similar masses but a very different structure.
Note, however, that this result depends on the energy gap used for the quark pairing. A large gap would make both cooling curves more similar.

\subsection{Including baryonic parity partners}
\label{parity}
With respect to hadronic models, there has been a number of studies looking at nuclear matter properties and also at stellar matter
with a different formulation of chiral symmetry and the baryonic fields. These refer to the so-called parity-doublet model
\cite{DeTar1989,Dexheimer2008a,Motohiro2015}.
The main feature of the doublet model is the inclusion of both, the positive and negative parity states of the baryons.
By arranging them in doublets, 
the components of the fields defining the parity partners $\varphi_+$ and $\varphi_-$
transform differently with respect to chiral transformations:
\begin{eqnarray}
&\varphi'_{+R}  =  R \varphi_{+R} & \varphi'_{+L}
= L \varphi_{+L} ~~,\ \nonumber \\ 
&\varphi'_{-R} = L \varphi_{-R} & \varphi_{-L}
= R \varphi_{-L} ~,
\end{eqnarray}
where $L$ and $R$ are rotations in the left- and right handed subspaces.
This behaviour is very similar to the mesonic analogue of the sigma meson and the pion as grouped together in the linear
sigma model. In the latter case,  the combination of sigma and pion  $\sigma^2 + \vec{\pi}^2$
is chirally invariant, whereas the separate terms are not. Here, including the opposite parity states helps to achieve the same
effect. It allows for a chirally invariant mass-like term in the
Lagrangian of the form:
\begin{eqnarray}
&m_{0}( \bar{\varphi}_- \gamma_{5} \varphi_+ - \bar{\varphi}_+
      \gamma_{5} \varphi_- ) =  \nonumber \\
&m_0 (\bar{\varphi}_{-L} \varphi_{+R} -
        \bar{\varphi}_{-R} \varphi_{+L} - \bar{\varphi}_{+L} \varphi_{-R} +
        \bar{\varphi}_{+R} \varphi_{-L}) ,
\end{eqnarray}
where $m_0$ is a mass parameter.

The SU(3) extension of the parity-doublet approach has been developed in Ref. \cite{Nemoto1998}. 
In general one can write the following SU(3)-invariant baryon-meson interaction
\begin{eqnarray}
{\cal L_B} &=& {\rm Tr} (\bar{\Xi} i {\partial\!\!\!/} \Xi)
+ m_0 {\rm Tr}( \left(\bar{\Xi} \gamma_5 \tau_2 \Xi \right) +
D^{(1)}_s {\rm Tr}( \bar{\Xi} \left\{ \Sigma, \Xi \right\} ) \nonumber \\ &+&
F^{(1)}_s {\rm Tr}( \bar{\Xi} \left[ \Sigma, \Xi \right] )
+S^{(1)}_s {\rm Tr}( \Sigma ) {\rm Tr}( \bar{\Xi} \Xi )
 \nonumber \\ &+&
 D^{(2)}_s {\rm Tr}( \bar{\Xi} \tau_3 \left\{ \Sigma, \Xi \right\} ) +
 F^{(2)}_s {\rm Tr}( \bar{\Xi} \tau_3 \left[ \Sigma, \Xi \right] )\nonumber \\ &+&
S^{(2)}_s {\rm Tr}( \Sigma ) {\rm Tr}( \bar{\Xi} \tau_3 \Xi ) +
D_v {\rm Tr}( \bar{\Xi} \gamma_\mu \left\{ V^\mu, \Xi \right\} ) \nonumber\\
&+&
F_v {\rm Tr}( \bar{\Xi} \gamma_\mu \left[ V^\mu, \Xi \right] ) +
 S_v {\rm Tr}( V^\mu) {\rm Tr}( \bar{\Xi} \gamma_\mu \Xi ) ~.
\label{lagrangian}
\end{eqnarray}
Here, $\Xi$ denotes the baryon octet that now also contains the doublets of all the octet states.  The $\tau$ matrices act on the doublet spinors. $\Sigma$ and $V^\mu$ represent the scalar and vector multiplets, respectively.
Note that there are now two sets of couplings $F^{(j)}, D^{(j)}, S^{(j)}$ with j=1,2 for the usual SU(3) invariant expressions.

After diagonalization in the doublet-spinor space, the following Lagrangian results
\begin{eqnarray}
{\cal L_B} &=& \sum_i (\bar{B_i} i {\partial\!\!\!/} B_i)
+ \sum_i  \left(\bar{B_i} m^*_i B_i \right) \nonumber \\ &+&
\sum_i  \left(\bar{B_i} \gamma_\mu (g_{\omega i} \omega^\mu +
g_{\rho i} \rho^\mu + g_{\phi i} \phi^\mu) B_i \right)~.
\label{lagrangian2}
\end{eqnarray}
The coupling constants for the baryons $B_i$ with the vector mesons $\omega$, $\rho$
and the strange meson $\phi$ are analogous to Eq. (\ref{vector}).
The main difference to the previous approaches appears in the expression for the effective baryon masses (neglecting the isovector scalar meson, for simplicity)
\begin{eqnarray}
m^*_i &=& \sqrt{ \left[ (g^{(1)}_{\sigma i} \sigma + g^{(1)}_{\zeta i}  \zeta
)^2 + (m_0+n_s m_s)^2 \right]}\nonumber \\
&\pm& g^{(2)}_{\sigma i} \sigma \pm g^{(2)}_{\zeta i} \zeta~, \label{mef}
\end{eqnarray}
where there are two sets of coupling constants to the baryons $i$ $g^{j}_{\sigma i}, g^{j}_{\sigma i}$, which are combinations of the original parameters. The signs $\pm$ distinguish the cases of opposite parity baryons.
As can be seen in the expression, in the case of vanishing scalar fields,
the various doublets are degenerate but not massless.
The expression contains a SU(3) breaking mass term with $m_s = 150$ MeV, responsible for the generation of an explicit mass corresponding to the number of strange quarks $n_s$ in each baryon.

\begin{figure}
\resizebox{0.5\textwidth}{!}{
  \includegraphics[width=7cm,height=7cm]{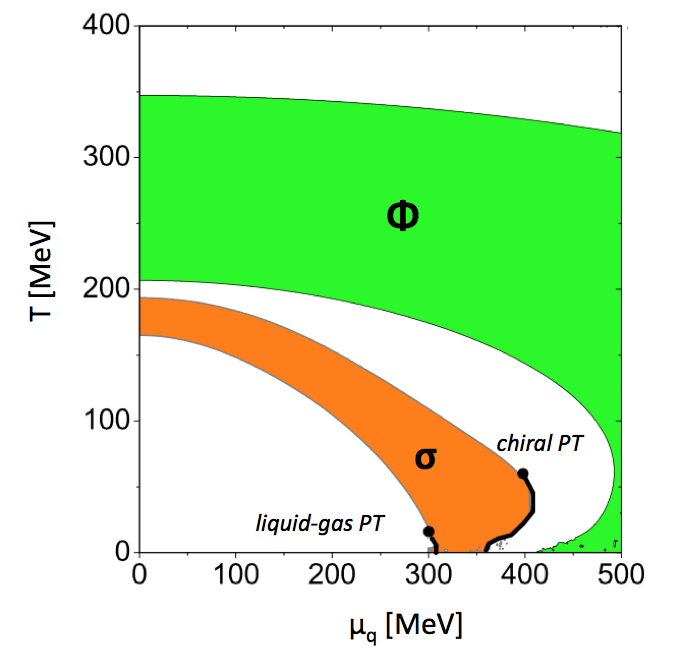}}

%\vspace{5cm}       % Give the correct figure height in cm
\caption{Phase diagram of the SU(3) parity-doublet model in the plane of temperature and quark chemical potential.
Two first-order phase transitions for isospin-symmetric matter can be observed, corresponding to liquid-gas and chiral transitions. The bands correspond to a change in value of  the $\Phi$ and $\sigma$ fields between 20 and 80 percent of their maximum value.}
\label{Tmuparity}       % Give a unique label
\end{figure}

With these relations at hand, we can perform a calculation analogous to the one in Section \ref{exvol} including quarks.
The resulting phase structure of the model for isospin-symmetric matter is given in Fig. \ref{Tmuparity}. As an interesting difference to the previous discussion, the model exhibits a second first-order phase transition, which is driven by the appearance of quarks as well as the parity-doublet partner of the nucleons. 
This transition ends in a critical end-point around a temperature of 60 MeV. Thus, it might be very useful to study the implications of such a phase structure in neutron star merger calculations that might sample the conditions close to such a critical point.

\begin{figure}
\resizebox{0.5\textwidth}{!}{
  \includegraphics[width=7cm,height=7cm]{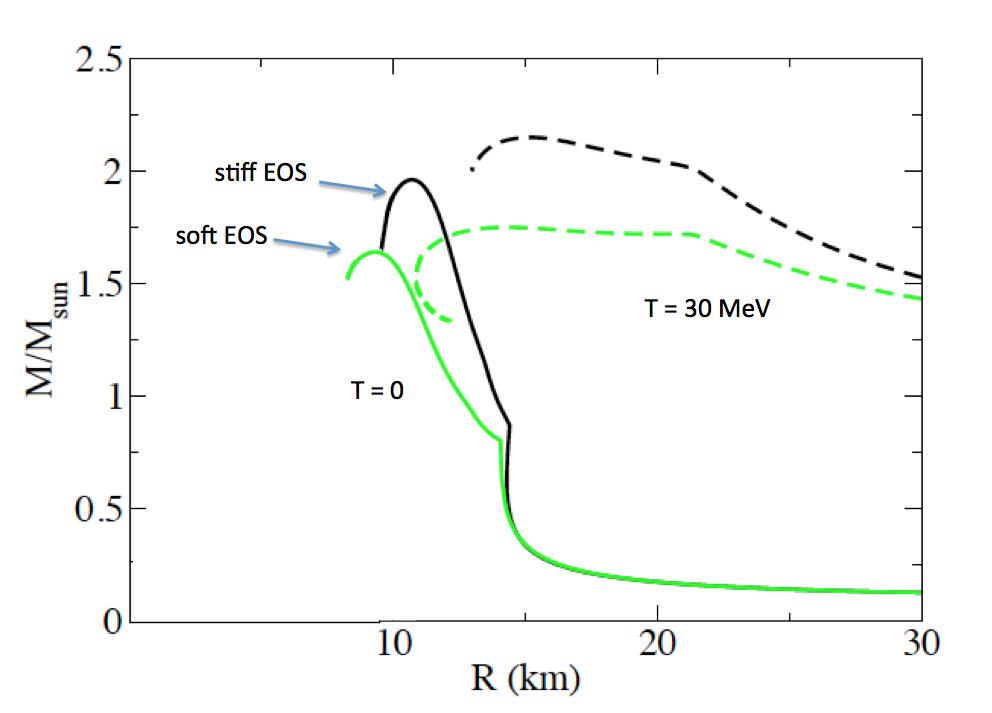}}
\caption{TOV solutions for  hybrid stars in the parity-doublet approach. Two different equations of state are compared. Results for vanishing temperature and $T = 30\, $MeV are shown.}
\label{MRparity}       % Give a unique label
\end{figure}

Besides the parametrization used in the calculation described above, which reproduces a high value for the compressibility at saturation, we define another parametrization with a  lower value for the compressibility. TOV solutions for both equations of state are shown in Fig. \ref{MRparity}. All curves for the star solutions show the familiar behaviour of a kink signalling the crossing of the first-order transition point, with a maximum between 1.7 and 2 $M_\odot$, respectively, for the soft and stiffer equation of state at zero temperature. In addition, the figure contains results for a fixed temperature of 30 MeV with slightly increased maximum masses. Note that the fixed temperature calculation is not an ideal approximation for a snapshot of a proto-neutron star thermal evolution, where a calculation at fixed entropy per baryon would have been more realistic and should be performed in future work. The temperature used mimics the maximum temperature effect to be expected in the core of the star.  It corresponds to an entropy per baryon of about 2 in the stellar center, which  is a reasonable early-stage value compared to numerical simulations \cite{Burrows1986,Pons1999,Pons2001}.  

\begin{figure}
\resizebox{0.5\textwidth}{!}{
  \includegraphics[width=7cm,height=7cm]{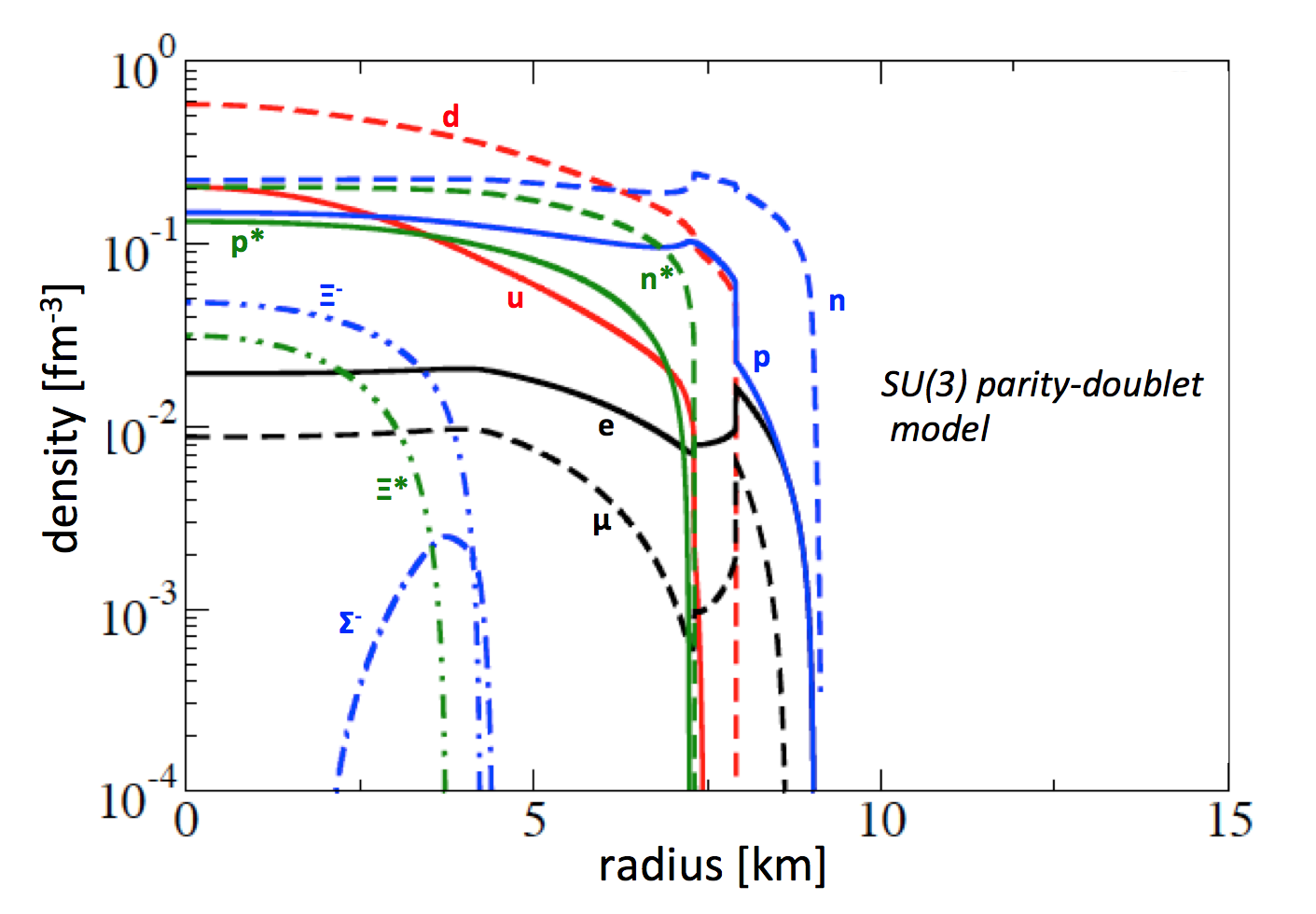}}
\caption{Particle densities as function of distance from the center of the star for the maximum mass star of the stiff equation of state of Fig. \ref{MRparity} at T=0. The complex
particle cocktail contains the octet of baryons and their parity doublets as well as quarks and leptons.}
\label{rhoparity}       % Give a unique label
\end{figure}

 The particle composition of these stars can be quite complex as is presented in Fig. \ref{rhoparity} for the stiff equation of state at zero temperature. These stars contain baryons, quarks, as well as the opposite parity states of the nucleons and some hyperons.

\begin{figure}
\resizebox{0.5\textwidth}{!}{
  \includegraphics[width=7cm,height=7cm]{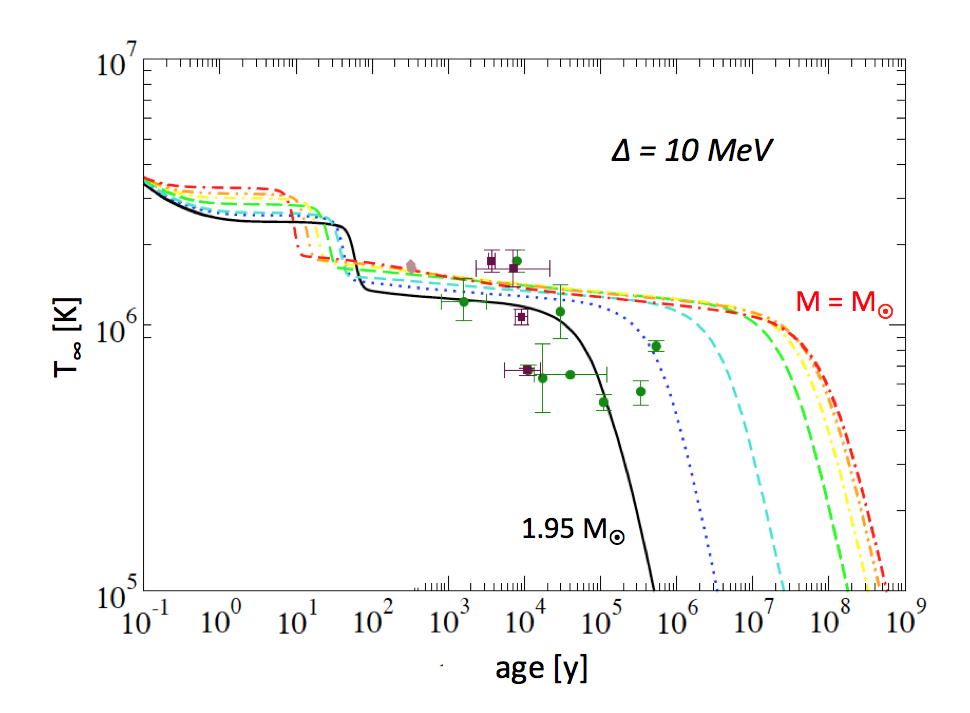}
 }
\vspace{.1cm}       % Give the correct figure height in cm
\caption{Stellar temperature as function of stellar life time resulting from a cooling simulation. Observational data are included as symbols.
The curves show results for different star masses. The simulation includes quark pairing with a gap of 10 MeV. }
\label{coolinggapparity}       % Give a unique label
\end{figure}

Finally, we look at the temperature evolution of stars within this model. Fig. \ref{coolinggapparity} shows results for a range of star masses for the stiff equation of state. 
The calculation includes a gap of 10 MeV for the quark pairing in the core. The figure also includes a range of observed stellar temperatures
and star ages derived from the spin-down of the pulsar as well as from kinematics, tracing the star back to the original supernova. 
Overall, there is a quite good agreement of the different measurements with the region of possible temperatures. However, a simultaneous 
measurement of star masses would be very helpful in constraining neutron star models via studies of stellar thermal evolution.

\section{Possible conflict with lattice results}
\label{lattice}

As has been documented in the previous sections, one nearly universal ingredient for all hybrid star calculations, within the framework discussed in the article as well as in many other calculations, is the inclusion of a strong repulsive force between the quarks \cite{Weissenborn2011a,Benic2014,Alford2005,Agrawal2010,Blaschke2010,Panda2004,Bonanno2012}. This force stiffens the equation of state and allows for the survival of a larger quark core without immediate collapse to a black hole. The calculation in Section \ref{model1} did not introduce a non-zero vector coupling of the quarks. However, as it was shown, this then only allows for a region of a mixed phase of quarks and baryons if one performs a Gibbs construction of the phase transition.

When one applies the different model approaches to matter at vanishing baryon chemical potential, one has the advantage to be able to compare
results with lattice QCD calculations. This has been used in adjusting the model parameters of the approaches, discussed in Sections \ref{model1}, \ref{exvol}, and \ref{parity}, and a direct comparison was shown in Fig. \ref{sigmaphi}. Overall, without too much fine-tuning a reasonable agreement with lattice data can be achieved for these star models.
Unfortunately, such comparison cannot easily go further, as lattice calculations at non-zero chemical potential are notoriously difficult. In this case,  a phase in the weighting factor of the 
path integral  leads to very problematic numerical cancellations in the sampling of the integral. Some effort has been put into a reweighting method to calculate observables at finite chemical potential. A main result was the computation of a critical end-point of a first-order transition line at T = 162 MeV and $\mu_B = 360\, $MeV with larger statistical errors. This result has been used in Section \ref{model1} to fix the model parameters.
However, it is still debated whether this critical point exists or not. Another way of extrapolating into the regime of non-zero chemical potential is to 
do a Taylor expansion of the grand canonical potential and calculate coefficients of an expansion in $\mu_B$.
The expansion of the pressure can be written as
\begin{eqnarray}
\frac{p(T,\mu_B)}{T^4}&=& \sum_{n=0}^{\infty}c_n(T)\left(\frac{\mu_B}{T}\right)^n ~\,\\
c_n(T)&=&\left. \frac{1}{n!} \frac{\partial^n(p(T,\mu_B)/T^4)}{\partial(\mu_B/T)^n}\right|_{\mu_B=0}~~.
\end{eqnarray}

\begin{figure}
\resizebox{0.5\textwidth}{!}{
  \includegraphics[width=7cm,height=6cm]{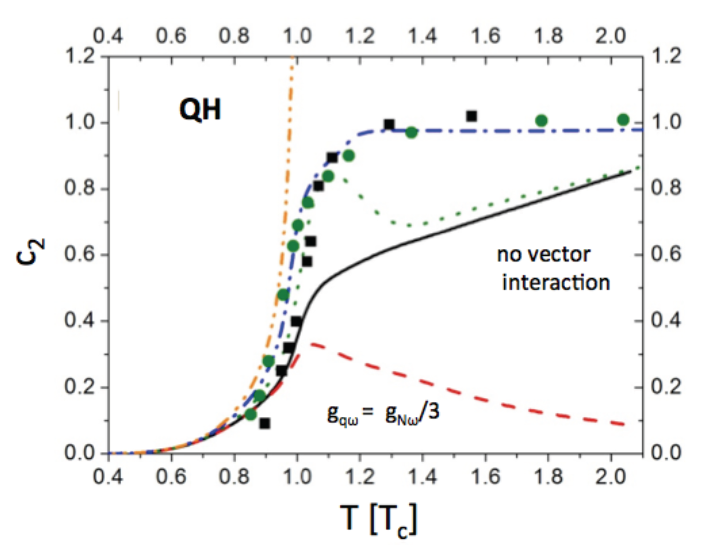}
}
%\vspace{5cm}       % Give the correct figure height in cm
\caption{Taylor coefficient of the pressure $c_2$ as function of temperature at zero chemical potential for the quark hadron model described in Section \ref{exvol}.
The solid line was obtained for zero interaction, the dashed line represents results for $g_{q\omega} = g_{N\omega}/3$. Again, a small interaction
strength causes results that are inconsistent with lattice data.}
\label{c2QH}       % Give a unique label
\end{figure}

\begin{figure}
\resizebox{0.5\textwidth}{!}{
  \includegraphics[width=7cm,height=6cm]{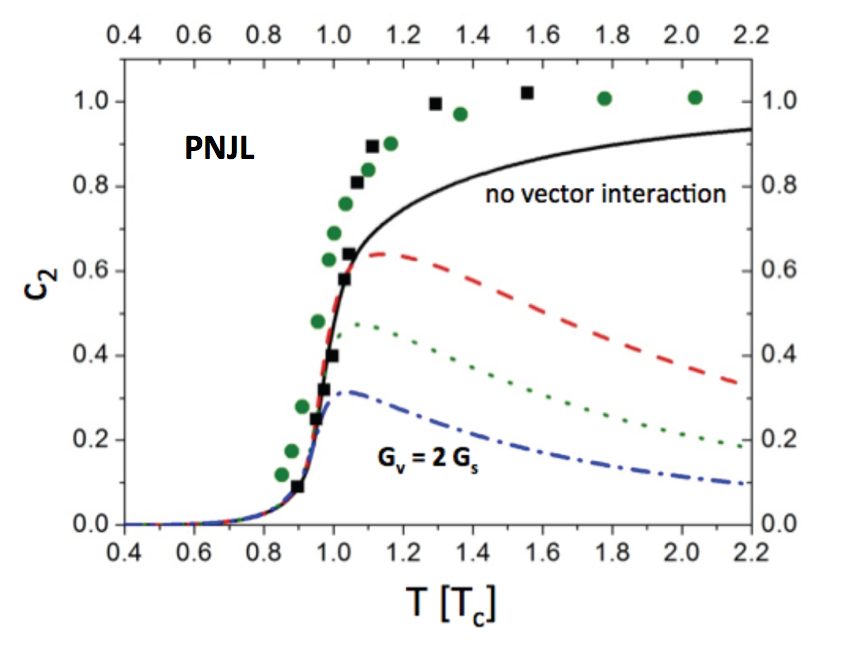}
}
%\vspace{5cm}       % Give the correct figure height in cmP
\caption{Same as Fig. \ref{c2QH} but for a standard PNJL model.
The curves correspond to different values of the repulsive quark-quark interaction, $G_v = 0, G_s/2, G_s, 2 G_s$, where $G_s$ is the strength of the scalar interaction. As can be seen, even a relatively small interaction
strength leads to strong disagreement with lattice data.}
\label{c2NJL}       % Give a unique label
\end{figure}

The advantage of this approach is that the coefficients $c_n$ can be calculated on the lattice as expectation values of operators at vanishing
chemical potential.
The first non-zero correction term of such a study is $c_2$.
Calculating the coefficient in the quark-hadron model discussed in Section \ref{exvol} and comparing it to lattice gauge simulations lead
to the results shown in \ref{c2QH} \cite{Steinheimer2011a}.
The comparison in the figure uses somewhat older lattice data \cite{Cheng2009}. For a more careful study with recent lattice data, see Ref. \cite{Steinheimer2014}.
Note, however, that all qualitative conclusions remain unchanged. Various theoretical curves are shown that correspond to different strengths of the vector interaction. As can be clearly seen, a non-zero vector coupling leads to very significant differences with respect to the lattice data. 
In order to demonstrate that this is not a model-specific result, an analogous calculation was done for the PNJL model.
These results,  depicted in Fig. \ref{c2NJL} demonstrate the same behaviour. A reasonably large value of the vector interaction coupling $G_V$ results in clear disagreement with the lattice simulations. Also, an analysis of the pseudo-critical phase transition line as function of chemical potential that, according to Ref. \cite{Bratovic2013}, signals a strong vector interaction term, does not change this picture. As it has been shown in Ref. \cite{Steinheimer2014}, the rather flat transition line is governed by the properties of  matter below the transition, i.e. the hadronic matter that naturally has a strong vector interaction, but not by the quark phase (see also Refs. \cite{Ferroni2011,Restrepo2015}). Therefore, this problem still persists, at least for the models currently under investigation.

%\begin{figure}
%\resizebox{0.5\textwidth}{!}{
%  \includegraphics[width=7cm,height=7cm]{parity_cooling.png}}
%\vspace{5cm}       % Give the correct figure height in cm
%\caption{Stellar temperature as function of life time, resulting from a cooling simulation. Observational data are included as symbols.
%The curves show results for different star masses.}
%\label{coolingparity}       % Give a unique label
%\end{figure}

\section{Conclusions}

In this article we have discussed several approaches connecting hadronic descriptions and quark models in a unifying way, which 
allows for first-order and higher-order phase transitions as well as crossovers. Such models 
can reproduce a mixed state of quarks and hadrons, even without the appearance of any sharp phase transition in dense matter.
As one interesting result, we observe twin star solutions for some model parameters, yielding two sets of hybrid stars with different strangeness content and radius but with the same mass. As was shown, these stars would have a distinctly different thermal evolution.

In accordance with other theoretical studies, we find that in order to obtain a star with a large quark core or a more extended mixed phase of quarks and hadrons, the quark interactions have to be strongly repulsive.
However, such assumption leads to problems in the agreement with lattice data at small densities (that study the Taylor coefficients for an expansion of the pressure with respect to baryon chemical potential). This problem should be addressed before one can make more definite claims about the feasibility of hybrid stars with large quark cores.  Hopefully, this point can be reasonably addressed without the need to render the strong interaction models too complex with the inclusion of many additional phenomenological terms.

\section*{Acknowledgements}
 SWS acknowledges support from the  Helmholtz International Center for FAIR. RN acknowledged financial support from CNPq and CAPES.

% BibTeX users please use
% \bibliographystyle{}
\bibliographystyle{prsty}
\bibliography{library.bib}
%\bibitem{RefJ} Author, Journal \textbf{Volume}, (year) page numbers.
%Author, \textit{Book title} (Publisher, place year) page numbers
% etc

\end{document}